\documentclass[a4paper,10pt]{article}
\usepackage{a4wide,makeidx}                             
\usepackage[english]{babel}                             
\usepackage{amsmath,amsfonts,verbatim,graphicx,float} 

\begin{document}

\begin{center}
{\Large \bf Momentum space trigonometric Rosen-Morse potential}
\end{center}

\vspace{0.5cm}
\begin{center}
C.\ B.\ Compean and M.\ Kirchbach \\
Instituto de Fisica, UASLP,
Av. Manuel Nava 6, Zona Universitaria,\\
San Luis Potosi, SLP 78290, M\'exico
\end{center}

\vspace{0.5cm}
\begin{flushleft}
{ We transform
 the trigonometric $S$ wave  Rosen-Morse potential
 to momentum space in employing its
property of being a harmonic angular function on the three-dimensional
hypersphere $S^3$.

    } 
\end{flushleft}

\vspace{0.5cm}
 \begin{flushleft} {PACS-key}
03.65.Ge, 02.30Uu
\end{flushleft}



\def\s{\mbox{\boldmath$\displaystyle\mathbf{\sigma}$}}
\def\J{\mbox{\boldmath$\displaystyle\mathbf{J}$}}
\def\K{\mbox{\boldmath$\displaystyle\mathbf{K}$}}
\def\P{\mbox{\boldmath$\displaystyle\mathbf{P}$}}
\def\p{\mbox{\boldmath$\displaystyle\mathbf{p}$}}
\def\hp{\mbox{\boldmath$\displaystyle\mathbf{\widehat{\p}}$}}
\def\x{\mbox{\boldmath$\displaystyle\mathbf{x}$}}
\def\0{\mbox{\boldmath$\displaystyle\mathbf{0}$}}
\def\bv{\mbox{\boldmath$\displaystyle\mathbf{\varphi}$}}
\def\hbv{\mbox{\boldmath$\displaystyle\mathbf{\widehat\varphi}$}}

\def\bg{\mbox{\boldmath$\displaystyle\mathbf{\gamma }$}}

\def\bl{\mbox{\boldmath$\displaystyle\mathbf{\lambda}$}}
\def\br{\mbox{\boldmath$\displaystyle\mathbf{\rho}$}}
\def\1{\mbox{\boldmath$\displaystyle\mathbf{1}$}}
\def\bfhh{\mbox{\boldmath$\displaystyle\mathbf{(1/2,0)\oplus(0,1/2)}\,\,$}}

\def\mn{\mbox{\boldmath$\displaystyle\mathbf{\nu}$}}
\def\amn{\mbox{\boldmath$\displaystyle\mathbf{\overline{\nu}}$}}

\def\mne{\mbox{\boldmath$\displaystyle\mathbf{\nu_e}$}}
\def\amne{\mbox{\boldmath$\displaystyle\mathbf{\overline{\nu}_e}$}}
\def\rlh{\mbox{\boldmath$\displaystyle\mathbf{\rightleftharpoons}$}}

\def\wm{\mbox{\boldmath$\displaystyle\mathbf{W^-}$}}
\def\hh{\mbox{\boldmath$\displaystyle\mathbf{(1/2,1/2)}$}}
\def\h00h{\mbox{\boldmath$\displaystyle\mathbf{(1/2,0)\oplus(0,1/2)}$}}
\def\znbb{\mbox{\boldmath$\displaystyle\mathbf{0\nu \beta\beta}$}}



\newcommand{\csch}{\textrm{ csch }}
\newcommand{\sech}{\textrm{ sech }}
\newcommand{\arccot}{\textrm{ arccot }}
\newcommand{\arccoth}{\textrm{ arccoth }}
\newcommand{\e}{\textrm { e}}

\newcommand{\be}{\begin{eqnarray}}
\newcommand{\ee}{\end{eqnarray}}
\newcommand{\nn}{\nonumber}

\vspace{1truecm}

\bigskip

Momentum space potentials obtained as
Fourier transforms  of central potentials 
are of interest in a variety of physics problems ranging from
condensed matter to particle physics. They can be viewed as instantaneous
propagators of the  fields
mediating the respective
interactions and are especially
important in Faddeev few-body calculations which
are more efficiently  carried out in momentum
than in position space.
Unfortunately, the power potentials of wide spread such as linear and harmonic
oscillator interactions
do not have well defined Fourier integrals \cite{Garz},
the inverse distance potential being the most prominent exception.

We here make the case that the $S$ wave  trigonometric Rosen-Morse potential
when considered as an angular function on the
three-dimensional (3D) surface of constant positive curvature,
the $S^3$ hypersphere,  allows for a momentum space transform 
that can be cast in closed form.

The  $\cot +\csc^2$ interaction, known as  the
trigonometric Rosen-Morse potential and  managed by SUSYQM 
\cite{Levai}  has in fact been
invented by Schr\"odinger in ref.~\cite{Schr40} and 
was originally introduced as 
an angular function on a three-dimensional (3D) surface of constant
positive curvature, the hypersphere $S^3$ embedded in a flat Euclidean space
of four dimensions, $E_4$.  
Up to additive constants it takes the form
\begin{equation}
V_{RM}(\chi )= -2B\cot\chi +\frac{\hbar^2}{2\mu d^2 } l(l+1)\csc^2\chi .
\label{RMI} 
\end{equation}
Here  $l$ is the value of the 3D angular momentum, $d$ is a matching 
length constant,
while  $\chi$ is the second polar angle in $E_4$.
In choosing the parameterization, $\chi =\frac{r}{R}$, for the angular 
variable,
where $R$ is the constant radius of $S^3$, while $r=R\chi $ 
is the length of the arc on the hyper-spherical surface, 
$V_{RM}(\chi )$ is usually  given the form of a potential in a
3D space, though  not a flat one.
The 3D flat Euclidean space, $E_3$, embedded in $E_4$ is described in
terms of a radius vector of an  absolute value,
$|{\mathbf r}|$, defined as $|{\mathbf r}|=R\sin \chi $.  Therefore,
the  $\chi $ parametrization corresponding to the correct 3D flat
space embedded in $E_4$ is, $\chi =\sin^{-1}\frac{|{\mathbf r}|}{R}$.

\noindent
The nature of the space, flat versus curved, is of minor importance
for the energy spectrum and the wave functions and reduces to the
interpretation of $R$. In flat space $R$ is viewed
as some matching length parameter, while  $S^3$  puts it 
on the firmer grounds of a parameter encoding the curvature.
Yet, regarding integral transformations such like
Fourier transforms to momentum space, the nature of the space
acquires significance through  the definition of the integration volume.
Trying to use flat space $E_3$ integral volume and a 3D plane wave to
Fourier transform $V_{RM}\left( \chi=\frac{r}{R}\right)$ 
as a function of the arc, $r$, is inconsistent
and  leads to a divergent Fourier integral.
In considering instead,  $V_{RM}$ 
as a function of the radius vector of the correct flat $E_3$ space, underlying
$E_4$,  allows for a Fourier transform that can be 
taken in closed form.    
Below we calculate the 4D Fourier transform of
$V_{RM}(\chi=\sin^{-1}\frac{|{\mathbf r}|}{R} )$ to momentum space.

\noindent
Angular potentials in extra dimensions are important because they
allow to replace complicated many-body problems in flat space
by effective two-body systems on curved spaces
with  the curvature parameter absorbing the many-body effects.
This is a well known technique which has been applied in several
physics problems ranging from plasma to instanton physics
\cite{plasma}-\cite{Nawa}.
Specifically the trigonometrical Rosen-Morse potential has found an  
interesting application in the physics of strongly interacting elementary
particles \cite{Quiry}. It  has been shown  to act as the exactly solvable 
extension to the quark confinement potential \cite{Cornell}
obtained in solving the equation of Quantum
Chromodynamics by the technique of simulations on a lattice, 
an observation reported in \cite{Quiry_2}. 
As another relevant application of same 
potential we wish to mention its employment  in the theory of quantum dots 
\cite{Kurochkin}.

Treating the interaction under discussion as an angular function on $S^3$
is possible because of its $SO(4)$ symmetry. The latter is best
understood in observing  that the Schr\"odinger equation with the
$\cot +\csc^2$ potential is closely related to  the
eigenvalue problem of the 4D angular momentum on $S^3$.
Through the paper we consider ordinary Euclidean flat space, $E_3$,   
embedded in a 4D Euclidean space, $E_4$, and parametrize   
the 3D spherical surface $S^3$  as $x_4^2+{\mathbf r}^2=R^2$ with 
$x_4=R\cos \chi$, and $|{\mathbf r}|=R\sin\chi$.
Here, $R$ is the constant hyper-radius of $S^3$.
The angular part,
$\widehat{ \underline{{\overline{\vert\,\, \vert}}}},$ 
of the 4D Laplace-Beltrami operator is proportional to
the operator of the squared 4D  angular momentum,
${\mathcal K}^2$, and given by,
\begin{eqnarray}
 \widehat{ \underline{{\overline{\vert\,\, \vert}}}} =  
\left[\frac{1}{\sin^2\chi }
\frac{\partial }{\partial \chi}
\sin^2\chi \frac{\partial }{\partial \chi } -
\frac{{\mathbf  L}^2 (\theta ,\varphi ) }{\sin^2 \chi }\right]
=-\kappa {\mathcal K}^2,&\quad&
 \kappa=\frac{1}{R^2}.
\label{lpls_4}
\end{eqnarray}
Here ${\mathbf L}^2(\theta ,\varphi )$ is the standard 3D orbital 
angular momentum operator in $E_3$,
the ordinary position space,\footnote{
The analogue on the 2D sphere, $S^2$, of constant radius
$|{\mathbf r}|=a$,  is the well known relation
${\vec \nabla}^2 =-\frac{1}{a^2}{\mathbf L}^2 $.
} $\chi $ is the second polar angle in $E_4$,
$\chi \in [0,\pi ]$, while $\kappa $ stands for the constant curvature.
 Consequently,  
the  Schr\"odinger  equation on $S^3$ becomes
\begin{eqnarray}
{\Big[} \frac{\hbar^2}{2\mu }\kappa {\mathcal K}^2 
&-& E(\kappa ){\Big]} \psi  (\chi, \kappa )=0,
\label{chi_eq}
\end{eqnarray}
where $\mu$ stands for the reduced mass.
The ${\mathcal K}^2$ eigenvalue-problem reads
\cite{Kim_Noz}
\begin{equation}
{\mathcal K}^2 \vert K l m \rangle = K(K+2)
\vert K l m \rangle, \quad 
\vert Klm\rangle \in \left( \frac{K}{2},\frac{K}{2} \right).
\label{Casimir_O4}
\end{equation}
The  $|Klm>$-levels belong to irreducible  
$SO(4)$ representations of the type  $\left(\frac{K}{2},\frac{K}{2} \right)$,
and the quantum numbers, 
$K$, $l$, and $m$ define the eigenvalues of the respective  
four--, three-- and two--dimensional angular momentum operators
upon the  state. These quantum numbers 
correspond to the $SO(4)\supset SO(3)\supset SO(2)$ reduction chain and
satisfy the branching rules,
$l=0,1,2,.. K$, and  $m=-l,..., +l$.
Multiplying eq.~(\ref{chi_eq}) by $(-\sin^2\chi )$ and changing variable to
$\psi  (\chi, \kappa  )=\sin \chi 
{\mathcal S} (\chi,\kappa )$,
results in the following Schr\"odinger equation,
\begin{eqnarray}
\left[ 
-\kappa \frac{\hbar^2}{2\mu }\frac{\mbox{d}^2 }{\mbox{d}\chi ^2}
+U_l(\chi ,\kappa) \right]
{\mathcal S} (\chi,\kappa  ) 
= E^{}(\kappa ) {\mathcal S} (\chi ,\kappa  ),&&\nonumber\\
U_l(\chi, \kappa)=\kappa \frac{\hbar^2}{2\mu } l(l+1)\csc^2\chi,&&
\label{chi_free}
\end{eqnarray} 
with  $U_l(\chi, \kappa)$ now having the meaning of
 centrifugal barrier on $S^3$.
As a different  reading to eqs.~({\ref{chi_eq}), and (\ref{chi_free})
one can say that
the  $\csc^2$ potential, in representing the
centrifugal barrier on the 3D hypersphere,  has $SO(4)$ as
potential algebra.
An important observation is that 
$SO(4)$ remains unaltered as potential algebra 
upon adding to the $\csc^2$ term
the harmonic function
\footnote{
Harmonic angular functions in $E_4$ are ${\mathcal K}^2$ 
eigenfunctions belonging to zero eigenvalues.
The  function $\cot\chi$ of the second polar angle is such a quantity, and
the counterpart on $S^3$ to the harmonic $S^2$ function,
$\ln \tan\frac{\theta }{2}$, of the first polar angle which
satisfies $\nabla^2\ln \tan \frac{\theta}{2}=0$.
The general mathematical theory of angular potentials and related harmonic 
functions
has been developed by Gabov in \cite{Gabov} and references therein.
} $\cot \chi $. 

This is visible from the fact
that the energy continues being a function of the
${\mathcal K}^2$ eigenvalues, $K(K+2)$ alone which
translate into the principle quantum number $n$ used in
\cite{Levai} as $n=K+1$. In effect, the $SO(4)$ symmetry of the
$\cot +\csc^2$ interaction allows to consider it as an angular function
on $S^3$, a circumstance that will substantially facilitate its transformation
to momentum space.
We here adopt the following parametrization of
the trigonometric Rosen-Morse potential as a function
of the second polar angle, $\chi$, on $S^3$, and the curvature:
\begin{eqnarray}
{\mathcal V}(\chi )= -2G\sqrt{{\kappa}}\cot \chi
+ {\kappa} \frac{\hbar^2}{2\mu } 
\frac{l(l+1)}{\sin^2 \chi }.
\label{crnl}
\end{eqnarray}
In Cartesian coordinates the $\cot \chi$ term equals 
$\frac{x_4}{|{\mathbf r}|}$, and stands in fact for two potentials
distinct by a sign and describing interactions on the
the respective Northern, and Southern hemi-spheres.
Correspondingly, their respective Fourier transforms 
to momentum space  become
\begin{eqnarray}
4\pi \Pi
( |\mathbf{q}| ) =-2G\sqrt{\kappa}\frac{2\mu}{\hbar^2}
\int_0^\infty d|x| |x|^3 \delta (|x| -R)\int_0^{2\pi} d \varphi 
\int_0^\pi 
d\theta \sin\theta 
\int_{0/\frac{\pi}{2}}^{\frac{\pi}{2}/\pi } d \chi \sin^2\chi 
e^{i|\mathbf{ q}|\frac{\sin\chi}{\sqrt{\kappa}}\cos\theta}
\cot \chi ,
\label{b2}
\end{eqnarray}
where the $\delta (|x| -R)$ function restricts $E_4$ to $S^3$.
Here, the 4D plane wave has been evaluated 
in reference  to a $z$ axis chosen along the
momentum vector (a choice justified in elastic scattering
\footnote{A consistent definition of the $E_4$ plane wave 
would require an Euclidean $q$ vector. However, for 
elastic scattering processes, of zero energy transfer, where $q_0=0$,
the $q$ vector can be chosen to lie entirely in $E_3$, 
and be identified  with the physical space-like momentum transfer.}), 
and a position vector of the confined particle having in general 
a non-zero  projection on the extra dimension axis in 
$E_4$, 
\begin{equation}
e^{i q\cdot  x}= 
e^{i|\mathbf{q}||\mathbf{ r}|\cos \theta }=
e^{i{|\mathbf{q}|}\frac{\sin \chi}{\sqrt{{\kappa}}} 
\cos \theta },
\quad |\mathbf{r}|=R\sin\chi=\frac{\sin \chi}{\sqrt{{\kappa}}}.
\label{4_PW_FF}
\end{equation}
On $S^3$  one has to distinguish between two types of momentum 
space potentials.
The first one, displayed in Fig.~\ref{primodos},
goes with $\chi \in {\Big[}0,\frac{\pi}{2}{\Big ]}$,  
corresponds to a positive $x_4$, and describes an increasing $|{\mathbf r}|$, 
while  the second refers
to $\chi \in {\Big[}\frac{\pi}{2},\pi {\Big ]}$, a negative $x_4$, and
describes a decreasing $|{\mathbf r}|$. 
The first type refers to the Northern 
hemisphere and reads
\begin{equation}
\Pi(|{\mathbf q}|)= 
c \frac{2\sin^2 
\frac{|{\mathbf q} |}{2
\hbar \sqrt{\kappa}}}{\left(\frac{|{\mathbf q}|}
{\hbar\sqrt{\kappa }}\right)^2}, \quad c=2G \frac{2\mu }{\hbar^2 {\kappa}}.
\label{prop_we}
\end{equation}
It is increasing  in the infrared,
finite at origin, and approaches asymptotically  the Coulomb 
propagator in the ultraviolet.
Such a type of behavior is required, for example,
in the description of confinement
phenomena  \cite{tereza}.
If one had treated instead the $\cot $ potential as a flat space interaction,
the 3D Fourier integral would have been divergent \cite{Quiry_2} 
and one would have been forced to introduce a $\pi $ range correlation 
function in order to get  it finite as we did in ref.~ \cite{SanCarlos}. 

In summary, one of the virtues of the curvature aspect of the $\cot$ 
interaction is that its $S^3$ Fourier transform comes out well defined.

\begin{figure}[b]
\center
\includegraphics[width=10.5cm]{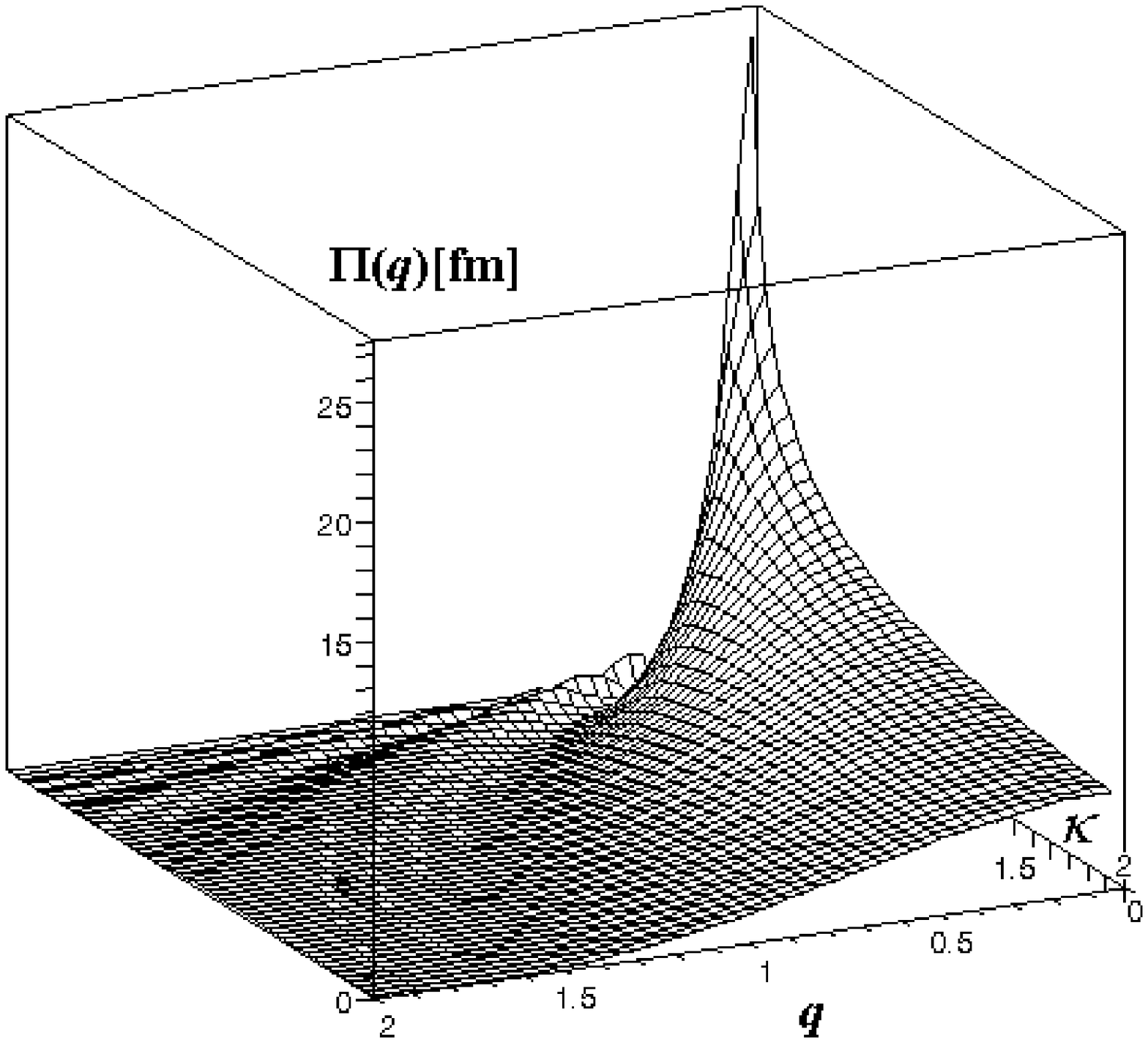}
\caption{ The curvature dependence of the momentum space
potential on the Northern 
hemisphere.
The Southern part appears mirrored with respect to
the horizontal plane.
Both potentials approach zero at infinity.
\label{primodos}}
\end{figure}

\vspace{0.11cm}

\noindent
{\bf Acknowledgments}\\

\noindent
Work supported by CONACyT-M\'{e}xico under grant number
CB-2006-01/61286.

\end{document}